\documentclass[useAMS,usenatbib,usedcolumn]{mn2e}
\def\HI{\ifmmode{\rm HI}\else{H\/{\sc i}}\fi}

\def\lsun{\ifmmode{{\mathrm L}_{\odot}}\else{L$_{\odot}$}\fi}

\def\msun{\ifmmode{{\mathrm M}_{\odot}}\else{M$_{\odot}$}\fi} 
\def\msunpc2{\ifmmode{{\mathrm M}_{\odot} \, {\mathrm{pc}}^{-2}}\else{M$_{\odot} \, {\mathrm {pc}}^{-2}$}\fi}

\def\kms{\ifmmode{{\mathrm{km \, s^{-1}}}}\else{${\mathrm{km \, s^{-1}}}$}\fi}

\def\la{\mathrel{\mathchoice {\vcenter{\offinterlineskip\halign{\hfil
$\displaystyle##$\hfil\cr<\cr\sim\cr}}}
{\vcenter{\offinterlineskip\halign{\hfil$\textstyle##$\hfil\cr
<\cr\sim\cr}}}
{\vcenter{\offinterlineskip\halign{\hfil$\scriptstyle##$\hfil\cr
<\cr\sim\cr}}}
{\vcenter{\offinterlineskip\halign{\hfil$\scriptscriptstyle##$\hfil\cr
<\cr\sim\cr}}}}}
\def\ga{\mathrel{\mathchoice {\vcenter{\offinterlineskip\halign{\hfil
$\displaystyle##$\hfil\cr>\cr\sim\cr}}}
{\vcenter{\offinterlineskip\halign{\hfil$\textstyle##$\hfil\cr
>\cr\sim\cr}}}
{\vcenter{\offinterlineskip\halign{\hfil$\scriptstyle##$\hfil\cr
>\cr\sim\cr}}}
{\vcenter{\offinterlineskip\halign{\hfil$\scriptscriptstyle##$\hfil\cr
>\cr\sim\cr}}}}}

\def\aj{AJ}%% Astronomical Journal
\def\apj{ApJ}%% Astrophysical Journal
\def\apjs{ApJS}%% Astrophysical Journal, Supplement
\def\aap{A\&A}%% Astronomy and Astrophysics
\def\mnras{MNRAS}%% Monthly Notices of the RAS
\def\pasp{PASP}%% Publications of the ASP
\usepackage[figuresright]{rotating}
\usepackage{lscape}
\usepackage{graphics}
\usepackage{graphicx}
\usepackage{epsfig}
\usepackage{multirow}
\usepackage{bigdelim}
\usepackage{bigstrut}
\usepackage{amsmath}
\usepackage{amssymb}

\defcitealias{Jaffe09}{J09}

\title[The colour-magnitude relation of Elliptical and Lenticular galaxies]
{The colour-magnitude relation of Elliptical and Lenticular galaxies in the ESO Distant Cluster Survey}
\author[Y.~L.~Jaff\'e et al.]{Yara L. Jaff\'e$^{1,2}$\thanks{E-mail: ppxyj@nottingham.ac.uk}, Alfonso Arag\'on-Salamanca$^1$, Gabriella De Lucia$^3$, Pascale Jablonka$^4$, \and Gregory Rudnick$^5$, Roberto Saglia$^{6,7}$ and Dennis Zaritsky$^8$\\ 
   $^1$School of Physics and Astronomy, University of Nottingham, University Park, Nottingham NG7 2RD, UK\\
   $^2$European Southern Observatory, Karl-Schwarzchild Strasse 2, 85748 Garching, Germany\\
   $^3$Osservatorio Astronomico di Trieste INAF, Via Tiepolo 11, 34143 Trieste, Italy\\
   $^4$Laboratoire d'Astrophysique, Ecole Polytechnique F\'ed\'erale de Lausanne (EPFL), CH-1290 Sauverny, Switzerland\\
   $^5$Department of Physics and Astronomy, University of Kansas, 1251 Wescoe Hall Dr., Lawrence, KS, 66045-7582, USA\\ 
   $^6$Max–Planck–Institut f¨ur extraterrestrische Physik, Giessenbachstrasse, 85748 Garching, Germany\\ 
   $^7$Universit¨atssternwarte, Scheinerstrasse 1, 81679 M¨unchen, Germany\\
   $^8$Steward Observatory, University of Arizona, 933 North Cherry Avenue, Tucson, AZ 85721, USA\\ }	

\begin{document}

\maketitle

\begin{abstract}
In this paper we study the colour-magnitude relation (CMR) for a sample of 172 morphologically-classified elliptical and S0 cluster
galaxies from the ESO Distant Cluster Survey (EDisCS) at $0.4\lesssim z \lesssim0.8$. The intrinsic colour scatter about the CMR is very small ($\langle \sigma_{\rm int} \rangle= 0.076$) in rest-frame $U-V$. However, there is a small minority of faint early-type galaxies (7\%) that are significantly bluer than the CMR. 
We observe no significant dependence of $\sigma_{\rm int}$ with redshift or cluster velocity dispersion. Because our sample is strictly morphologically-selected, this implies that by the time cluster elliptical and S0 galaxies achieve their morphology, the vast majority have already joined the red sequence. The only exception seems to be the very small fraction of faint blue early-types. Assuming that the intrinsic colour scatter is due to differences in stellar population ages, 
we estimate the galaxy formation redshift $z_{\rm F}$ of each cluster and find that  $z_{\rm F}$  does not depend on the cluster velocity dispersion. 
However, $z_{\rm F}$ increases weakly with cluster redshift within the EDisCS sample. This trend becomes very clear when higher redshift clusters from the literature are included. 
This suggests that, at any given redshift, in order to have a population of fully-formed ellipticals and S0s they needed to have formed most of their stars $\simeq2$--$4\,$Gyr prior to observation. That does not mean that \textit{all} early-type galaxies in \textit{all} clusters formed at these high redshifts. It means that the ones we see already having early-type morphologies also have reasonably-old stellar populations. This is partly a manifestation of the ``progenitor bias'', but also a consequence of the fact that the vast majority of the early-type galaxies in clusters (in particular the massive galaxies) were already red (i.e., already had old stellar populations) 
by the time they achieved their morphology. Elliptical and S0 galaxies exhibit very similar colour scatter, implying similar stellar population ages. The scarcity of blue S0s indicates that, if they are the descendants of spirals whose star-formation has ceased, the parent galaxies were already red when they became S0s. This suggests the red spirals found preferentially in dense environments could be the progenitors of these S0s.
We also find that fainter early-type galaxies finished forming their stars later (i.e., have smaller $z_{\rm F}$), consistent with the cluster red sequence being built over time and the brightest galaxies reaching the red sequence earlier than fainter ones. Combining the CMR scatter analysis with the observed evolution in the CMR zero point we find that the early-type cluster galaxy population must have had their star formation truncated/stopped over an extended period $\Delta t \gtrsim 1\,$Gyr.

\end{abstract}


%%%%%%%%%%%%%%%%%%%%%%%%%%%%%%%%%%%%%%%%%%%%%%%%%%%%%%%%%%%%%%%%%%%%%%%%%%%%%%%
%                                                                             %
%  1. Introduction                                                            %
%  \label{sec:introduction}                                                   %
%                                                                             %
%%%%%%%%%%%%%%%%%%%%%%%%%%%%%%%%%%%%%%%%%%%%%%%%%%%%%%%%%%%%%%%%%%%%%%%%%%%%%%%
\section{Introduction}
\label{sec:introduction}

Galaxy clusters have proven to be very useful laboratories for the study of galaxy formation and evolution. They can provide large and diverse galaxy samples across practically small areas of sky. Although the relative importance of nature and nurture in shaping galaxy evolution remains debated, it is well established that many galaxy properties in the nearby Universe correlate strongly with their environment. In particular, 
\citet{PostmanGeller1984} and \citet{Dressler80} found that the morphological mix changes as a function of local galaxy density, with early-type galaxies (E/S0) dominating in locally-dense environments, while late types (spirals and irregular galaxies) become increasingly important towards lower densities. Moreover, at low redshift there is an obvious bimodality in the colours of cluster galaxies that is well correlated with  
morphology \citep[e.g. see recent work by][]{Conselice06,Wang07}. The ``blue cloud'' is dominated by spirals and irregular galaxies, and the prominent ridge of red galaxies (the so-called ``red sequence'') is mainly populated by E/S0 galaxies. This implies that morphology correlates at some level with the stellar population content (and hence colour).

\citet[][]{Baum1959}, \citet[][]{Faber1973} and \citet[][]{Caldwell1983}  established the existence of a red sequence of cluster ellipticals in the local Universe and showed that these galaxies have systematically redder colours with increasing luminosity. \citet[][]{VS1977} and \citet[][]{SV1978a, SV1978b} found that this colour-magnitude relation (hereafter CMR) is universal.  The detailed study of the $UVK$ colours of local cluster early-type galaxies carried out by \citet[][]{BLE} confirmed Sandage \& Visvanathan's anticipation that both S0s and ellipticals follow the same relation. Furthermore, they also showed that the observed scatter about the CMR is very small ($\sim 0.04$ mag in
$U-V$ for their sample). 

In the past decade, a number of studies have shown that the CMR of elliptical galaxies holds at progressively higher redshifts, at least up to $z=1.4$ \citep[e.g.][]{Ellis97, Stansford1998, vanDokkum2000, vanDokkum2001, Blakeslee03, Mei06b, Lidman2008}. As a consequence, the CMR is, arguably, one of the most powerful scaling relations obeyed by  the early-type galaxy population at the cores of clusters, encoding important information about their formation history.

The slope of the CMR has traditionally been interpreted as the direct consequence of a mass-metallicity relation \citep[e.g.][]{Faber1973,Larson1974,Gallazzi2006}. The classical explanation of this mass-metallicity sequence is based on the idea that star-formation-induced galactic outflows would be more efficient at expelling metal-enriched gas in low-mass galaxies than in massive ones \citep[e.g.][]{Larson1974, DekelSirk1986, Tremonti2004, DeLucia04a, KobayashiSpringelWhite2007, FinlatorDave2008, ArimotoYoshii1987}. An alternative interpretation in which the CMR is predominantly an age sequence would imply that the relation changes significantly with redshift as less massive galaxies approach their formation epochs. The possibility that age is the main driver for the CMR was ruled out by observations of clusters at intermediate redshift that showed that the slope of the CMR evolves little with redshift \citep{KodamaArimoto, Kodama1998}. Nevertheless, weak age trends along the CMR have been claimed \citep[e.g.][]{Ferreras1999,Poggianti20001,Nelan2005}, even though it seems clear that they are not the main physical driver.

\citet[][]{BLE} interpreted the small scatter about the CMR as the result of small age differences at a given galaxy mass. The tightness of the relation then implies very synchronized star-formation histories for these galaxies. Larger colour scatter would imply later episodes of star-formation, or a wider range in galaxy formation redshifts. These results are not only indicative of the passive evolution of elliptical galaxies but also of an early formation epoch \citep[$z > 2$--$3$; e.g.][]{BLE,Blakeslee03,Mei09}. Studies of absorption-line indices in the spectra of early-type galaxies also imply old ages \citep[e.g.][]{Trager1998}. Some evidence has been found that the mean stellar ages of early-type galaxies may depend on their stellar mass in the sense that lower-mass galaxies appear to have formed their stars at later epochs than the more massive ones \citep[e.g.][]{Thomas2005}, although \citet{Trager2008} find no such trend in their study of the Coma cluster.

The above interpretation of the nature of the CMR, although traditionally accepted, has an important problem:  it assumes that all red-sequence galaxies that we see today can be identified as red-sequence members of high redshift galaxy clusters. As noted by \citet{vanDokkum1996}, this assumption is probably wrong because of the so-called \textit{progenitor bias}: if the progenitors of some early-type galaxies were spirals at higher redshift, they would not be included in the higher redshift samples, which biases the studied population towards older ages. This effect has been corroborated by recent studies of the CMR evolution. \citet{DeLucia07} found a significant deficit of faint red cluster galaxies at $0.4\lesssim z \lesssim 0.8$ compared to galaxy clusters in the local Universe. They conclude that the red sequence population of high redshift clusters does not contain all the progenitors of nearby red sequence cluster galaxies \citep[see also][and references therein]{DeLucia2009}. \citet{Tanaka08} also find such deficit in a galaxy cluster at $z = 1.1$. We will come back to this issue in Sections~\ref{sec:results} and~\ref{sec:conclusions}.

In this paper, we study the CMR for a total sample of 174 morphologically-selected elliptical and S0 galaxies contained in 13 galaxy clusters and groups at $0.4 \lesssim z \lesssim 0.8$ from the ESO Distant Cluster Survey \citep[EDisCS]{white}, taking advantage of the availability of extensive Hubble Space Telescope (HST) imaging obtained with the Advanced Camera for Surveys (ACS) and extensive ground-based imaging and spectroscopy (see Section~\ref{sec:data}). We interpret the scatter about the CMR as a proxy for age (or formation time $t_{F}$), and study  its dependence on intrinsic galaxy properties such as their luminosities and morphologies, and the role of the environment as quantified by the mass/velocity dispersion of the clusters. We complement the scatter analysis with information derived from the zeropoint of each cluster's CMR to constrain not only the formation epoch of early-type galaxies but also the duration of their formation phase.

\begin{figure*}
\begin{center}
  \includegraphics[scale=0.71,angle=270]{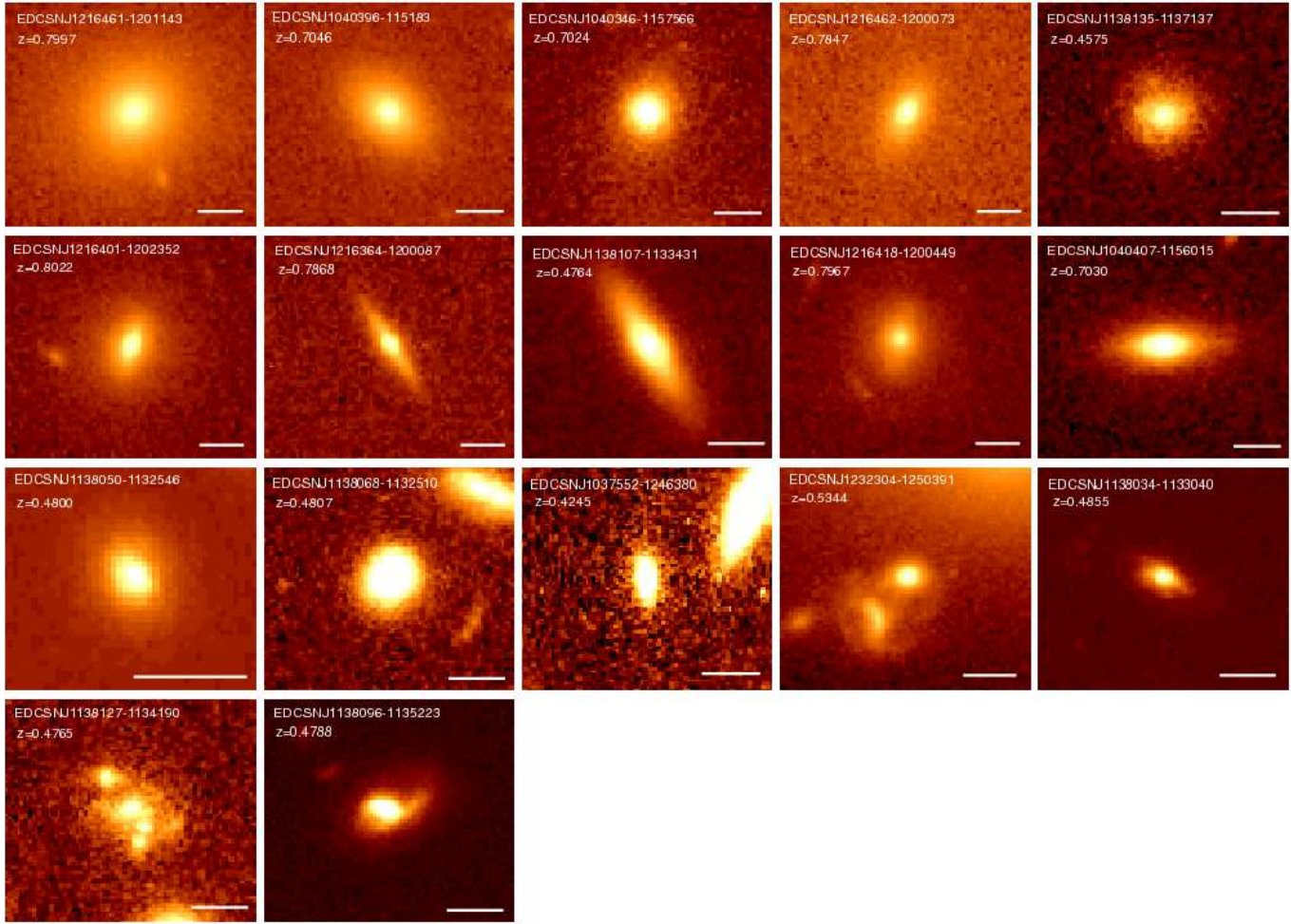}
\end{center} 
\caption{HST F814W images for a representative sample of ellipticals (top row), S0s (second row) and ``blue tail'' galaxies (third row).  
The last row shows the 2 galaxies that were excluded from our study due to morphological misclassification. 
The rightmost 2 galaxies in the third row (blue tail) exhibit some degree of perturbation in their morphologies (see text for further discussion).
The horizontal white lines correspond to $5\,$kpc.}
 \label{mosaic}
\end{figure*}

It is important to point out that, even though there is much evidence suggesting that the CMR scatter 
is principally driven by galaxy age \citep[e.g.][]{KodamaArimoto, KauffmannCharlot1998, Bernardi05}, metallicity variations 
could also contribute to it \citep{Nelan2005}. If that is the case, the stellar ages
we derive in this paper would be lower limits since the colour scatter we measure would contain both an age component and a metallicity one. 
Taking the results of \citet{Nelan2005} at face value, the maximum scatter in metallicity at a given velocity dispersion (luminosity) is $\sim0.1$dex, 
implying that at most we could be systematically underestimating the stellar ages by $\sim0.15$dex, where we have assumed \citet{Worthey94} $3/2$ age-metallicity degeneracy law. Notwithstanding this caveat, even if the absolute ages were affected at this level, it is not unreasonable to expect that the effect on relative ages (the main focus of this paper) would be smaller.

This paper is organized as follows: our dataset and the sample of early-type galaxies are described in Section~\ref{sec:data}. In Section \ref{sec:method}, we explain the ``scatter-age method'' used in this work to study the formation histories of early-type galaxies. Section~\ref{sec:results} shows the scatter measured about the CMRs, Section~\ref{sec:SFH} the results of the scatter-age test and Section~\ref{sec:zeropoint} the zeropoint analysis. Finally, we discuss our results and draw conclusions in Section~\ref{sec:conclusions}.

Throughout this work, we use Vega magnitudes and adopt a WMAP cosmology \citep[][]{Spergel07} with $\Omega_{\rm M}=0.28$,  $\Omega_{\Lambda}=0.72$ and $H_{0}=70\,$Km$\,$s$^{-1}$Mpc$^{-1}$.

%***********************************************************************************

\section{The Data}
\label{sec:data}

\begin{table*}
\begin{center}
 \caption{Properties of the cluster sample. The columns correspond to: cluster ID, spectroscopic redshift, line-of-sight 
 velocity dispersion, number ($N$) of early-type galaxies (E+S0), number ellipticals ($N$(E)) taken into account in the scatter calculation, and 
number of ``blue tail'' members ($N_{\rm blue}$). See text for details. Cluster redshifts and velocity dispersions were taken from \citet{Halliday2004} and \citet{MJ2008}.}.
\label{cluster_sample}
\begin{tabular}{lccccc}
\hline
Cluster name & $z$ & $\sigma_{v}\,$(km/s) & $N$(E+S0) & $N$(E) & $N_{\rm blue}$\\
\hline\\[-2mm]
Cl1037.9$-$1243a &0.4252	&$537^{+46}_{-48}$	 &12	&7 	&1	\\[1mm]
Cl1138.2$-$1133a &0.4548	&$542^{+63}_{-71}$	 &9		&7	&1	\\[1mm]
Cl1138.2$-$1133	 &0.4796	&$732^{+72}_{-76}$	 &16	&11	&3	\\[1mm]
Cl1232.5$-$1250	 &0.5414	&$1080^{+119}_{-89}$ &23	&15	&3	\\[1mm]
Cl1354.2$-$1230a &0.5952	&$433^{+95}_{-104}$	 &4	  &4	&0	\\[1mm]
Cl1103.7$-$1245a &0.6261	&$336^{+36}_{-40}$	 &4	  &4	&0	\\[1mm]
Cl1227.9$-$1138	 &0.6357	&$574^{+72}_{-75}$	 &8	  &3	&0	\\[1mm]
Cl1054.4$-$1146	 &0.6972	&$589^{+78}_{-70}$	 &16	&14	&2	\\[1mm]
Cl1040.7$-$1155	 &0.7043	&$418^{+55}_{-46}$	 &8	  &7	&0	\\[1mm]
Cl1054.7$-$1245a &0.7305	&$182^{+58}_{-69}$	 &6	  &5	&0	\\[1mm]	
Cl1054.7$-$1245	 &0.7498	&$504^{+113}_{-65}$	 &18	&13	&0	\\[1mm]
Cl1354.2$-$1230	 &0.7620	&$648^{+105}_{-110}$ &5	  &4	&2	\\[1mm]
Cl1216.8$-$1201	 &0.7943	&$1018^{+73}_{-77}$	 &31	&18	&0	\\[1mm]
\hline 
\end{tabular} 
\\ 
\end{center}
\end{table*}

EDisCS originated as an ESO Large Programme designed to study cluster structure and cluster galaxy evolution over a large fraction of cosmic time
\citep{white}. The complete dataset consists of 20 fields containing galaxy clusters at redshifts between 0.4 and 1. The sample was constructed selecting 30 of the highest surface brightness candidates in the Las Campanas Distant Cluster Survey \citep{Gonzalez2001}, and confirming the presence of an apparent cluster and a possible red sequence with VLT 20-min exposures in two filters. From these candidates, 10 of the highest surface brightness clusters were followed up in each of two bins at estimated redshifts $0.45 < z_{\rm est} < 0.55$ and $0.75 < z_{\rm est} < 0.85$, where $z_{\rm est}$ was based on the magnitude of the putative brightest cluster galaxy. 

For the 20 fields with confirmed cluster candidates, deep optical imaging was taken using the FORS2 camera and spectrograph at the Very Large Telescope (VLT). A full description of the available optical images and photometry is given in \citet{white}. In brief, the optical photometry consists of $B$, $V$ and $I$ imaging for the 10 intermediate redshift ($z_{\rm est}\simeq 0.5$) cluster candidates and $V$, $R$ and $I$ imaging for the 10 high redshift ($z_{\rm est}\simeq 0.75$) cluster candidates. Typically, the integration times were 45min for the intermediate redshift sample and 2h for the high redshift sample. In addition, near-IR $J$ and $K$ photometry was obtained using the SOFI camera at the New Technology Telescope (NTT) (Arag\'on-Salamanca et al., in preparation). This was obtained for all but three fields, two of them were rejected due to bad weather and an additional field was rejected because its spectroscopic redshift histogram was not cluster-like. FORS2 at the VLT was also used to obtain multi-object spectroscopy for $\sim100$ objects in each field \citep[see][]{Halliday2004, MJ2008}. The redshift data were used to confirm the physical reality of the clusters and measure their velocity dispersions. This process revealed that several of the fields contained multiple clusters and groups at different redshifts \citep{Halliday2004, MJ2008, Poggianti2006}.  Additional follow-up includes narrow-band H$\alpha$ imaging \citep{Finn2005} and XMM X-ray observations \citep{Johnson2006} for a subset of the clusters. Moreover, ten of the highest redshift clusters in the sample were also imaged with HST's ACS camera in the F814W filter, from which reliable visual morphologies were derived for a large sample of cluster and field galaxies \citep[see][for details]{morph}. 

For consistency with our previous work on the CMR of EDisCS cluster galaxies \citep{DeLucia07}, in the following we use magnitudes and colours measured in seeing-matched images with $FWHM=0.8\,$arcsec (the typical seeing in the optical images), using a fixed $1.0\,$arcsec radius circular aperture. This aperture represents a compromise between minimizing sky-subtraction and contamination errors and being as close as possible to measuring global colours. We note that $1.0\,$arcsec corresponds to $\simeq7\,$kpc at $z\simeq0.7$, and at these redshifts early-type galaxies in the luminosity range considered here have half-light radii $\sim3\,$kpc \citep{Treu2005, Trujillo2007}. The associated photometric errors were derived by placing empty apertures in regions of the image without detected objects to estimate accurately the sky contribution to the error budget. This is justified since  the sky noise is the dominant source of error when measuring the aperture magnitudes of our faint galaxies \citep[see][for details]{white}, 

In this paper, we focus on a sub-sample of the EDisCS dataset consisting of 174 galaxies in 13 clusters and groups. This selection was based on the following criteria:
\begin{enumerate} 
 \item The galaxies must be spectroscopically-confirmed cluster/group members \citep{Halliday2004, MJ2008}. This ensures a very clean sample , avoiding the uncertainties introduced by other cluster membership criteria such as photometric redshifts \citep{Rudnick09, Pello09}. The penalty is, of course, a significant reduction in the sample size.
 \item They must have early-type morphology (E or S0), based on visual classification from HST images \citep{morph}.
We note that by imposing this, the sample reduces to galaxies within the 10 fields observed with HST, 
 \item The galaxies should belong to clusters/groups with at least 4 early-type members in order to measure the CMR scatter with reasonable accuracy. This allows us to detect the presence of an intrinsic colour scatter with $>3\sigma$ confidence in all cases.
\end{enumerate}
Since all spectroscopically-confirmed members in the HST-covered area have been morphologically classified, the selection function of our sample is the same as that of the spectroscopic sample. This means that for all practical purposes our early-type galaxy sample behaves like the original $I$-band selected spectroscopic sample  \citep{MJ2008}, and therefore as a rest-frame $B$-band selected sample. On average we typically reach $M_B<-18.5$, with some cluster-to-cluster variation. Although all our analysis has been carried out using this empirically-defined $I$-band selected sample, 
thus maximizing the sample size, we have checked that using a rest-frame $B$-band luminosity selected sample would not have altered any of our conclusions. 
We have also checked that the spectroscopic selection function  does not affect our conclusions. We produced Monte Carlo realizations of the colour-magnitude diagram of our sample taking into account the empirical selection function determined by \citet{MJ2008} and found that the simulated colour scatter agrees very well
with the measured one. 

To test the robustness of the morphologies for the galaxies in our sample, we re-examined visually their HST images (see Figure~\ref{mosaic}) 
and found only two galaxies that had been misclassified as early-types in \citet{morph}. The first one, EDCSNJ1138096$-$1135223, is clearly not an elliptical and shows a very perturbed morphology. 
The second one, EDCSNJ1138127$-$1134190, is in a dense group of (probably) interacting galaxies and there is a brighter elliptical very close to the position of this object. It is obvious that the wrong galaxy was classified.  The last two lines of Table~\ref{outsiders} show some of the properties of these two galaxies and their HST F814W images are shown in the bottom row of Figure~\ref{mosaic}. It is not surprising that both galaxies are substantially bluer than the red-sequence. They are also quite faint, where visual classification is, perforce, less reliable. These two misclassified galaxies were removed from our sample and will not be discussed further. We noticed that 6 out of the remaining 172 galaxies (3.5\%) have signs of perturbation, although their early-type morphology is clear. Interestingly, 2 out of these 6 slightly perturbed galaxies are significantly bluer than the CMR. We will come back to this in Section~\ref{sec:results}.

The cluster sample with the corresponding redshifts, line-of-sight velocity dispersions ($\sigma_{v}$) and number of early-type members 
is shown in Table \ref{cluster_sample}. The $\sigma_{v}$'s are taken from \citet[][]{Halliday2004} and \citet[][]{MJ2008}. The reliability of these velocity dispersions as mass estimators has been confirmed by weak lensing \citep{Clowe2006} and X-ray \citep{Johnson2006} estimates. 

%***********************************************************************************

\section{Method: the scatter-age test}
\label{sec:method}

The scatter-age test carried out in this paper was developed by \cite{BLE} as a reasonably simple, yet powerful method for constraining the formation history of early-type galaxies. They applied it at $z\sim0$ to galaxies in the Virgo and Coma clusters. \citet{Ellis97} applied the same test to galaxies in three $z\sim0.54$ clusters. We apply it here to a much larger cluster sample, covering a significant cluster mass range. We have the added advantage that since 1997 the uncertainty in the cosmological parameters, and thus the transformation of redshift into look-back time, has decreased considerably. We also benefit from a large and homogeneous galaxy sample in a wide range of redshift and cluster velocity dispersion (or cluster mass). This method has been used by many authors in the past \citep[e.g.][]{vanDokkum1998,Stansford1998,Blakeslee03,Blakeslee2006,Mei06b,Mei09,hilton09}. A description of the specific steps we took to perform the scatter-age test follows.

\begin{figure}
\begin{center}
\includegraphics[width=0.5\textwidth]{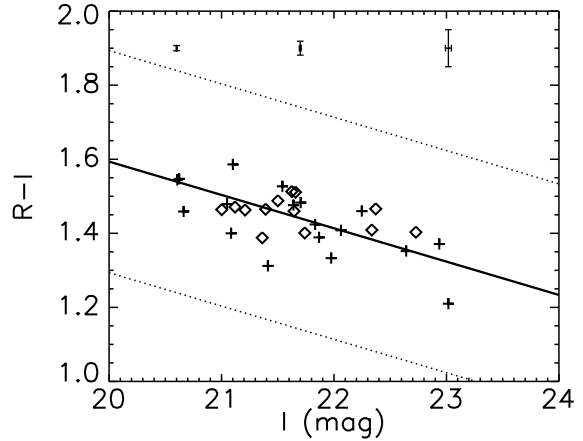}
\end{center} 
\caption{The $R-I$ vs. $I$ colour-magnitude diagram for the early-type galaxies in cluster CL1216.8$-$1201 at $z=0.79$ is shown here as an example. Elliptical galaxies are represented by ``+'' signs, and S0s by open diamonds. The solid line shows a linear fit to the colour-magnitude relation with the slope value determined by \citet{DeLucia07}. See text for details. The dotted lines correspond to $\pm0.3\,$mag from the CMR. For reference, the typical sizes of the error bars are plotted on the top of the figure as a function of magnitude.}
 \label{cmdsamplefig}
\end{figure}

We first constructed colour-magnitude diagrams (CMDs) of the early-type galaxies in each cluster using the photometric bands closest to rest-frame $U$ and $V$. Rest-frame $U-V$ measures the strength of the $4000$\AA\ break and is therefore a very age-sensitive broadband colour (see Section \ref{subsec:col} for a detailed justification of our choice of observed colour). For most of the clusters this choice required CMDs of $R-I$ versus $I$, but  for the three lowest redshift clusters we used  $V-I$ versus $I$ (see Table \ref{individual_scatters}). We then fitted a linear CMR for each cluster using a fixed slope of $-0.09$ \footnote{This value was previously used in \citet{DeLucia07} to construct V-I CMRs for colour selected red-sequence galaxies from the EDisCS database} and setting the zeropoint to the median colour. This procedure is very robust, in particular for groups and clusters with small numbers of galaxies where the CMR slope cannot be determined to sufficient accuracy. Our results are not sensitive to the exact choice of slope since in general the CMR is reasonably flat and redshift-independent \citep{Holden04}. As an example, the CMD for the early-type galaxies in cluster CL1216.8$-$1201 is shown in Figure~\ref{cmdsamplefig}.

For each cluster, the observed scatter in the galaxy colours about the CMR ($\sigma_{\rm obs}$) was computed as the r.m.s. of the residuals (in the colour direction) between the observed colours and the fitted CMR. We reject outliers in this process by imposing the condition that galaxy colours should be within $\pm0.3\,$mag from the CMR. While other methods such as the biweight scatter estimator used by other authors \citep[e.g.][]{Mei09} reject outliers implicitly, we chose to do it explicitly. We discuss these outliers in some detail later. 

Following \citet{BLE}, the intrinsic scatter ($\sigma_{\rm int}$) was then obtained by subtracting, in quadrature, the mean value of the photometric colour uncertainty from the observed scatter. The colour uncertainties range from $0.012$ to $0.021$ \citep{white}, and have therefore little effect on the observed scatter (see Section~\ref{sec:results}).   

\begin{figure}
\begin{center}
\includegraphics*[width=0.5\textwidth]{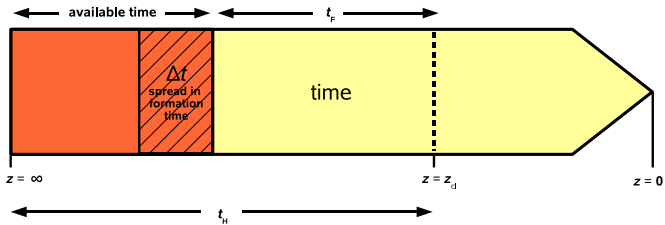}\vspace{-3.5cm}
\end{center} 
\caption{Time-line of the Universe illustrating the different parameters used in Equations~\ref{scatter_eq} and~ \ref{beta_eq}. The time arrow starts on the left at the beginning of the universe ($z=\infty$). The orange region shows the total available time galaxies can use to form stars.  $\Delta t$ is the time galaxies actually spend forming stars. From then on, the cluster galaxies are assumed to evolve passively until the observed redshift ($z_{cl}$). We define $t_{\rm F}$ as the time elapsed from the epoch when star formation ended until the cosmic time corresponding to $z_{cl}$.
}
 \label{bigarrowoftime}
\end{figure}

\cite{BLE} used $\sigma_{\rm int}$ to constrain the formation history of the galaxies by assuming the following relationship between the colour scatter and the colour evolution of the stellar population \citep{BLE}:
\begin{equation}
\delta(U-V)_0=\dfrac{d(U-V)_0}{dt}(t_{\rm H}-t_{\rm F})\beta\leq\sigma_{\rm int},
\label{scatter_eq}
\end{equation} 
where $t_{\rm H}$ is the age of the universe at the cluster redshift, $t_{\rm F}$ is the look-back time from then to the epoch at which star formation ended, and $d(U-V)_0/dt$ (where the subscript ``0'' denotes rest-frame) is derived from galaxy evolution models. This factor should be reasonably well understood as it is mainly governed by main-sequence evolution (for a given IMF). In this equation, $\beta$ parameterizes the spread in formation time $\Delta t$ as a fraction of the total available time:
\begin{equation}
\beta=\dfrac{\Delta t}{(t_{\rm H}-t_{\rm F})}.
\label{beta_eq}
\end{equation} 
Thus, $\beta=1$ implies no synchronization, i.e. the galaxies in the sample formed using all the available time, while $\beta=0.1$ would mean a high degree of synchronization, with all the galaxies forming in the last 10\% of the available time.
Figure \ref{bigarrowoftime} illustrates the time-line defined by Equations~\ref{scatter_eq} and~\ref{beta_eq}.

We calculated the $d(U-V)_0/dt$ factor using \citet[hereafter BC03]{bc03} models\footnote{For the range of ages discussed in this paper and the optical colours on which we base our conclusions, using alternative population synthesis models such as those of \citet{Maraston2005} would not change our results.}
for a passively-evolving stellar populations that formed in a single burst of $0.1\,$Gyr duration. The exact burst duration does not have a significant effect on our conclusions provided that it is much shorter than $t_{\rm F}$. Solar metallicity ($Z_\odot=0.02$), a \citet{Chabrier2003} initial mass function (IMF) and no dust attenuation were assumed.  Using alternative IMFs \citep[e.g.][]{Salpeter55} would not alter our conclusions because for the stellar masses of interest (given the range in $t_{\rm F}$), the relative differences in the IMFs are only minor \citep[cf.\ ][]{BLE}. The use of models with solar metallicity is partially motivated by the results presented in \citet{SB09}, where ages and metallicities are derived for EDisCS red-sequence galaxies from their absorption line indices. \citet{SB09} found solar-metallicity models agreed well with the observed galaxy spectra.  We discuss the effect of assuming different metallicities later.

\begin{figure}
\begin{center}
  \includegraphics[width=0.5\textwidth]{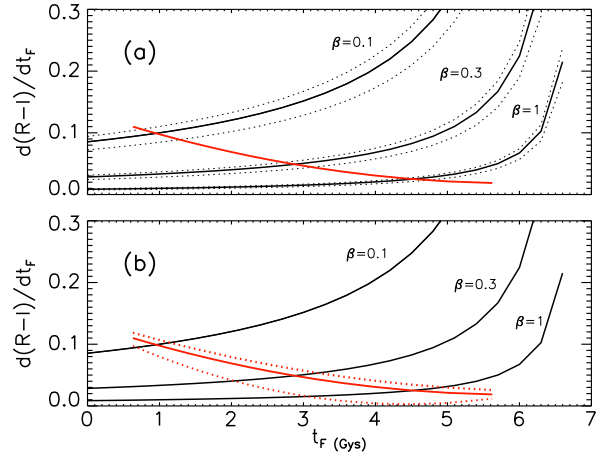}
  \end{center} 
\caption{Illustration of the colour-scatter method and the associated random (a) and systematic (b) uncertainties for the cluster CL1216.8$-$1201 at $z=0.79$. 
The rate of colour change ($d{\rm Colour}/dt$) as a function of $t_{\rm F}$ is shown by the identical solid red lines in panels~(a) and~(b). 
In this example we use the observed $(R-I)$, very close to rest-frame $(U-V)$ at the cluster redshift, computed using 
BC03 stellar population models for galaxies with solar metallicity and a single star formation burst of $0.1\,$Gyr duration (see text for details). 
The scatter about the CMR provides an upper limit to the allowed rate of colour evolution,   
parametrized by Equation~\ref{line_eq} for a given $\beta$ (Equation~\ref{beta_eq}). 
This constraint is shown in both panels by the identical solid black lines, as derived  
from the intrinsic colour scatter for this cluster and assuming three different values of $\beta$.
The intersection between the observational lines (solid black) and the model ones (solid red) constrains $t_{\rm F}$. 
The dotted black lines in panel~(a) correspond to the $\pm1\sigma$ random errors affecting the solid black line as a result of the 
observational uncertainty in the colour scatter. 
The red dotted lines in panel~(b) illustrate the effects of systematic model uncertainties (e.g. metallicity) on $t_{\rm F}$. These   lines correspond to models with the same star formation history as for the red solid line but different metallicities: the upper line has $Z_{\rm sub-solar}=0.008$, 
whilst the lower line has $Z_{\rm super-solar}=0.05$.}
 \label{errorsources}
\end{figure}

Figure~\ref{errorsources} shows how Equation~\ref{scatter_eq} can be used to constrain the star-formation history of the galaxies from the colour scatter $\sigma_{\rm int}$ of our richest cluster. It also illustrates the effect of the relevant random and systematic uncertainties. In Figure~\ref{errorsources}(a), the red solid line represents $d(R-I)/dt$ (very close to rest-frame $d(U-V)_0/dt$) as a function of $t_{\rm F}$, calculated from the BC03 models. The black solid lines are defined by the equation 
\begin{equation}
\dfrac{d(R-I)}{dt}=\dfrac{\sigma_{\rm int}}{(t_{\rm H}-t_{\rm F})\beta}
\label{line_eq}
\end{equation} 
for several values of $\beta$. Equation~\ref{scatter_eq} implies that the allowed region lies below the black lines, and thus the intersection of the red and black lines provides a constraint (upper limit) on $t_{\rm F}$ for a given $\beta$. It is clear that the colour scatter alone cannot be used to constrain $t_{\rm F}$ and $\beta$ simultaneously. The dotted black lines in Figure~\ref{errorsources}(a) represent the $\pm1\sigma$ random errors\footnote{The errors in the colour scatter were estimated from the 16\% and 84\% confidence levels in the $\chi^2$  distribution of the measured $\sigma_{\rm int}$, which correspond to $\pm1\sigma$ uncertainties.}  in the colour scatter, showing their effect on the $t_{\rm F}$ uncertainty. The effect of systematic uncertainties, such as changing the model chemical composition, are shown on panel~(b). The dotted red lines correspond to stellar population models with non-solar metallicity ($Z_{\rm sub-solar}=0.008$ and $Z_{\rm super-solar}=0.05$). It is immediately apparent that the effect of systematic model uncertainties such as these in the calculation of $t_{\rm F}$ is, in general, significantly larger than that of photometric random errors. This implies that absolute values of $t_{\rm F}$ must be interpreted with great caution. However, it is not unreasonable to assume that these systematics would affect all the galaxies similarly, making any {\it differential\/} or {\it comparative\/} study precise and, hopefully, robust. Unless otherwise stated, we consider random uncertainties only when discussing $t_{\rm F}$ since our study is largely comparative, but it is important to bear in mind that substantial systematic uncertainties do exist.

\subsection{Colour dependence of the derived $t_{\rm F}$}
\label{subsec:col}

\begin{figure}
\begin{center}
  \includegraphics[width=0.5\textwidth]{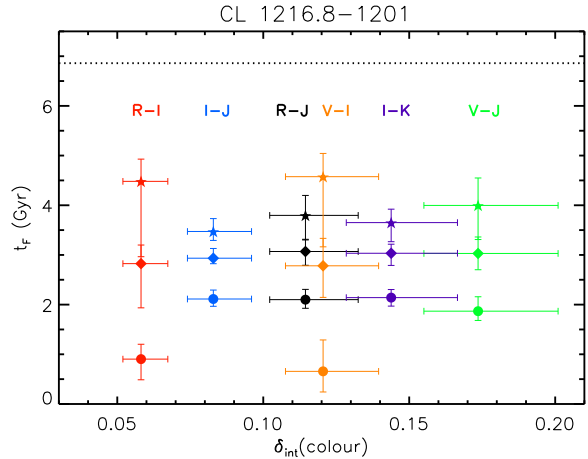}
\end{center} 
\caption{The derived formation times of the early-type galaxies in the cluster CL1216.9$-$1201 at $z=0.79$ are plotted against the intrinsic CMR scatter measured for the different colours used in the scatter analysis. With the exception of $I-J$ and $I-K$ all colours straddle the $4000$\AA\ break at the cluster redshift. Filled stars correspond to $\beta=1.0$, diamonds to $\beta=0.3$ and circles to $\beta=0.1$ (see Equation~\ref{beta_eq}). The black dotted line indicates the age of the Universe ($t_{\rm H}$) at the cluster redshift. In this Figure the errors in $t_{\rm F}$ are calculated as the sum (in quadrature) of the random and systematic uncertainties discussed in the text. Within the errors, most colours yield consistent values of $t_{\rm F}$ (particularly for $\beta \geq 0.3$). However, for $\beta=0.1$ the optical-optical and optical-infrared colours give discrepant results possibly due to the difficulty in modeling the contribution of asymptotic giant branch stars.}
 \label{allcolours}
\end{figure}

Previous studies have noted that colours which bracket the $4000$\AA\ break provide the most sensitive indicators of age changes, yet the least affected by photometric errors \citep[e.g.][]{Blakeslee2006}. For this reason, and following \citet{BLE}, we decided to carry out the age-scatter test using observed colours close to rest-frame $U-V$. Nevertheless, it is instructive to study how the actual colour choice could affect our results. We used our richest cluster, CL1216.8$-$1201, as the ideal test bed for this purpose. Using the galaxy sample in this cluster, we carried out the scatter-age test with CMDs compiled for different sets of colours. Most colours straddle the $4000$\AA\ break at $z=0.79$ (the cluster redshift), with the exceptions of $I-J$ and $I-K$. The values of $\sigma_{\rm int}$ and $t_{\rm F}$ derived for each colour are plotted in Figure~\ref{allcolours} for different values of $\beta$. It is clear that, at least for $\beta\geq0.3$, all the colours provide a consistent value of $t_{\rm F}$ within the combined random and systematic errors. However, for $\beta=0.1$ the optical-optical and optical-infrared colours give discrepant results. This is probably because at the relevant stellar population ages ($1$--$2\,$Gyr) asymptotic giant branch stars have a potentially large, and very uncertain, contribution to the near-infrared galaxy emission, making the model predictions very unreliable \citep{Maraston2005, ConroyGunn2009}. The colour with the smallest scatter and smallest scatter uncertainty is $R-I$, which is the closest to rest-frame $U-V$ at this redshift. This provides additional justification for the use of observed colours that are the closest match to rest-frame $U-V$ in our analysis.

\begin{table*}
\begin{center}
 \caption{Values of the measured intrinsic scatter ($\sigma_{\rm int}$) and the calculated formation time ($t_{\rm F}$) and redshift 
($z_{\rm F}$) for solar metallicity and $\beta=0.1, 0.3$ and $1.0$. The associated uncertainties correspond to the random errors (16\% and 84\% confidence levels in a $\chi^2$  distribution). Recall that $t_{\rm F}$ is the lookback time, from the cluster redshift, since star formation ceased.
\label{individual_scatters}}
\begin{tabular}{lccccccccc}
\hline
Cluster name & $z$ 	& CMR 	&$\sigma_{\rm int}$ 	& $t_{\rm F}(\beta=0.1)$ & $t_{\rm F}(\beta=0.3)$& $t_{\rm F}(\beta=1.0)$	\qquad\qquad &$z_{\rm F}(\beta=0.1)$	&$z_{\rm F}(\beta=0.3)$	&$z_{\rm F}(\beta=1.0)$\\
             &     	& colour      	&                   	&  (Gyr)                 & (Gyr)                  & (Gyr) 			&	&	&	\\
\hline\\[-2mm]
Cl1037.9$-$1243a&0.4252	&$V-I$	&$0.09^{+0.03}_{-0.01}$	   &$2.3^{+0.4}_{-0.6}$	&$4.4^{+0.3}_{-0.4}$	&$6.0^{+0.2}_{-0.3}$	&$0.76^{+0.08}_{-0.1}$	&$1.3\pm0.1$	&$2.1^{+0.1}_{-0.2}$\\[1mm]
Cl1138.2$-$1133a&0.4548	&$V-I$	&$0.10^{+0.03}_{-0.02}$    &$2.4^{+0.2}_{-0.3}$	&$3.34^{+0.1}_{-0.2}$	&$4.0^{+0.1}_{-0.1}$	&$0.82^{+0.04}_{-0.06}$	&$1.03^{+0.03}_{-0.05}$	&$1.22\pm0.03$\\[1mm]
Cl1138.2$-$1133	&0.4796	&$V-I$	&$0.09^{+0.03}_{-0.01}$	   &$2.1^{+0.3}_{-0.4}$	&$3.7^{+0.2}_{-0.3}$  	&$5.1^{+0.2}_{-0.3}$	&$0.79^{+0.06}_{-0.07}$	&$1.19^{+0.07}_{-0.09}$	&$1.8\pm0.1$\\[1mm]
Cl1232.8$-$1201	&0.5414	&$R-I$	&$0.08^{+0.02}_{-0.01}$	   &$1.8^{+0.3}_{-0.4}$	&$3.8^{+0.2}_{-0.3}$  	&$5.1^{+0.2}_{-0.3}$	&$0.82^{+0.06}_{-0.07}$	&$1.40^{+0.008}_{-0.1}$	&$2.6\pm0.2$\\[1mm]
Cl1354.2$-$1230a&0.5952	&$R-I$	&$0.04^{+0.03}_{-0.009}$   &$2.2^{+0.5}_{-1}$	&$4.2^{+0.4}_{-1}$	&$5.7^{+0.3}_{-0.7}$	&$1.0^{+0.1}_{-0.3}$	&$1.7^{+0.2}_{-0.4}$	&$2.8^{+0.3}_{-0.6}$\\[1mm]
Cl1103.7$-$1245a&0.6261	&$R-I$	&$0.08^{+0.07}_{-0.02}$	   &$1.0^{+0.6}_{-0.9}$	&$3.1^{+0.4}_{-1}$	&$4.7^{+0.3}_{-0.8}$	&$0.8^{+0.1}_{-0.2}$	&$1.4^{+0.2}_{-0.4}$	&$2.2^{+0.2}_{-0.5}$\\[1mm]
Cl1227.9$-$1138	&0.6357	&$R-I$	&$0.09^{+0.04}_{-0.02}$	   &$0.5^{+0.4}_{-0.4}$	&$2.5^{+0.3}_{-0.6}$	&$4.1^{+0.2}_{-0.4}$	&$0.71^{+0.07}_{-0.06}$	&$1.2^{+0.1}_{-0.2}$	&$1.8^{+0.1}_{-0.2}$\\[1mm]
Cl1054.4$-$1146	&0.6972	&$R-I$	&$0.06^{+0.02}_{-0.01}$	   &$1.0^{+0.3}_{-0.5}$	&$2.9^{+0.2}_{-0.3}$	&$4.5^{+0.2}_{-0.3}$	&$0.86^{+0.07}_{-0.1}$	&$1.44^{+0.09}_{-0.1}$	&$2.4\pm0.2$\\[1mm]
Cl1040.7$-$1155	&0.7043	&$R-I$	&$0.05^{+0.02}_{-0.01}$	   &$1.1^{+0.4}_{-0.7}$	&$3.2^{+0.3}_{-0.6}$	&$4.9^{+0.2}_{-0.4}$	&$0.9\pm0.1$	&$1.6\pm0.2$	&$2.7^{+0.2}_{-0.3}$\\[1mm]
Cl1054.7$-$1245a&0.7305	&$R-I$	&$0.11^{+0.06}_{-0.02}$	   &$0.5^{+0.5}_{-0.4}$	&$1.6^{+0.5}_{-1}$	&$4.0^{+0.2}_{-0.5}$	&$0.80^{+0.1}_{-0.07}$	&$1.4^{+0.1}_{-0.2}$	&$2.1^{+0.1}_{-0.3}$\\[1mm]	
Cl1054.7$-$1245	&0.7498	&$R-I$	&$0.10^{+0.02}_{-0.01}$    &$0.6^{+0.3}_{-0.4}$	&$1.8^{+0.3}_{-0.4}$	&$3.6^{+0.1}_{-0.2}$	&$0.84^{+0.06}_{-0.08}$	&$1.30^{+0.08}_{-0.11}$	&$1.90^{+0.06}_{-0.1}$\\[1mm]
Cl1354.2$-$1230	&0.7620	&$R-I$	&$0.05^{+0.03}_{-0.01}$	   &$0.9^{+0.5}_{-0.8}$	&$2.9^{+0.4}_{-0.9}$	&$4.6^{+0.3}_{-0.7}$	&$0.9^{+0.1}_{-0.2}$	&$1.6^{+0.2}_{-0.4}$	&$2.7^{+0.3}_{-0.5}$\\[1mm]
Cl1216.8$-$1201	&0.7943	&$R-I$	&$0.058^{+0.009}_{-0.006}$ &$0.9^{+0.2}_{-0.3}$	&$2.8^{+0.2}_{-0.2}$	&$4.5^{+0.1}_{-0.2}$	&$0.96^{+0.05}_{-0.07}$	&$1.6\pm0.1$	&$2.75^{+0.09}_{-0.2}$\\[1mm]
\hline 
\end{tabular}
\end{center} 
\end{table*}

\begin{table*}
\begin{center}
 \caption{Main characteristics of the comparison samples.}
 \label{comparison_samples}
\begin{tabular}{llllcclcc}
\hline
      & Cluster name 	&  $z$	& $\sigma_{\rm int}$ &Colour used & Ref.&  $\sigma_{v}$  & Ref. & Symbol in \\
	& 	&  	&  & in $\sigma_{\rm int}$& for $\sigma_{\rm int}$&(Km/s) & for $\sigma_{v}$ &Fig.~\ref{scatter_z}   \\
\hline\\[-2mm]
	&Coma	&0.0231	&$0.056\pm0.01$		&$U-V$	&1	&$821^{+49}_{-38}$ 	&	&$\divideontimes$	\\[1mm]	
	&Virgo	&0.0038	&$0.044\pm0.01$		&$U-V$	&1	&$632^{+41}_{-29}$	&	&$\divideontimes$	\\[1mm]
	&CL1358+62	&0.3283	&$0.079\pm0.01$		&$B-V$	&2	&$1027^{+51}_{-45}$	&	&$\divideontimes$	\\[1mm]
low-$z$	&CL10412-65	&0.510	&$0.131\pm0.027$		&$(U-V)_{z=0}$	&3	&$681^{+256}_{-185}$	&	&$\divideontimes$	\\[1mm]
	&CL10016+16	&0.546	&$0.06\pm0.01$		&$(U-V)_{z=0}$	&3	&$1127^{+166}_{-112}$	&7       &$\divideontimes$	\\[1mm]
	&CL10054-27	&0.563	&$0.06\pm0.01$		&$(U-V)_{z=0}$	&3	&$230\pm18$	&		&$\divideontimes$	\\[1mm]
\hline\\[-2mm]
	&MS 1054-0321	&0.831	&$0.070\pm0.008$ 	&$(U-B)_{z=0}$	&4	&$1156\pm82$		&4 &$\lozenge$		\\[1mm]
	&RX J0152.7-1357&0.834	&$0.050\pm0.005$ &$(U-B)_{z=0}$	&4	&$1203^{+96}_{-123}$	&4	&$\lozenge$		\\[1mm]
	&CL1604+4304	&0.897	&$0.031\pm0.003$ &$(U-B)_{z=0}$	&4	&$703\pm110$	&4	&$\lozenge$		\\[1mm]
high-$z$	&CL1604+4321	&0.924	&$0.043\pm0.006$ &$(U-B)_{z=0}$		&4	& $582\pm167$	&4	&$\lozenge$		\\[1mm]
	&RDCS J0910+5422	&1.106	&$0.060\pm0.009$	&$(U-B)_{z=0}$		&4	&	$675\pm190$&4	&$\lozenge$		\\[1mm]
	&RDCS J1252.9-2927	&1.237	&$0.112\pm0.022$	&$(U-B)_{z=0}$		&4	& $747^{+74}_{-84}$	&4	&$\lozenge$		\\[1mm]
	&RX J0849+4452	&1.261	&$0.070\pm0.014$	&$(U-B)_{z=0}$		&4	& $740^{+113}_{-134}$	&4	&$\lozenge$		\\[1mm]
	&RX J0848+4453	&1.270	&$0.049\pm0.027$	&$(U-B)_{z=0}$		&4	& $650\pm170$	&4	&$\lozenge$		\\[1mm]
\hline\\[-2mm]
very high-$z$&XMMU J2235.3-2557	&1.39	&	$0.055\pm0.018$&$J-Ks$		&5	&	$762\pm265$ &	&$\square$	\\[1mm]
	&XMMXCS J2215.9-1738	&1.46	&$0.12\pm0.05$	&$z_{850}-J$	&6 	&	$580\pm140$& &$\bigtriangleup$	\\[1mm]
\hline
\end{tabular}
\end{center}

\begin{flushleft}

Note: Following \citet{Blakeslee2006}, the scatter in $(U-B)_{z=0}$  was transformed into $(U-V)_{z=0}$ scatter by adding $0.04$. 

References:  (1) \citet{BLE}; (2) \citet{vanDokkum1998}; (3) \citet{Ellis97}; (4) \citet{Mei09} and references therein; (5) \citet{Lidman2008}; (6) \citet{hilton09}; (7) \citet{Borgani99}

\end{flushleft}
\end{table*}

%***********************************************************************************
\section{The scatter in the colour-magnitude relation}
\label{sec:results}

\subsection{The CMR scatter for different clusters}
\label{subsec:scatter}

For each cluster, we calculated the intrinsic colour scatter $\sigma_{\rm int}$ following the method described in Section \ref{sec:method}. The values of $\sigma_{\rm int}$ and the actual colours used for each cluster are listed in Table~\ref{individual_scatters}. At this stage, we exclude from this calculation all galaxies whose colours are $>0.3\,$mag bluer than the fitted CMR. The number of galaxies excluded in each cluster is listed in Table~\ref{cluster_sample}. In total, there are 12 (7\%) of these blue early-type galaxies in our sample. We justify this approach and discuss these galaxies later. 

\begin{figure}
\begin{center}
  \includegraphics[width=0.5\textwidth]{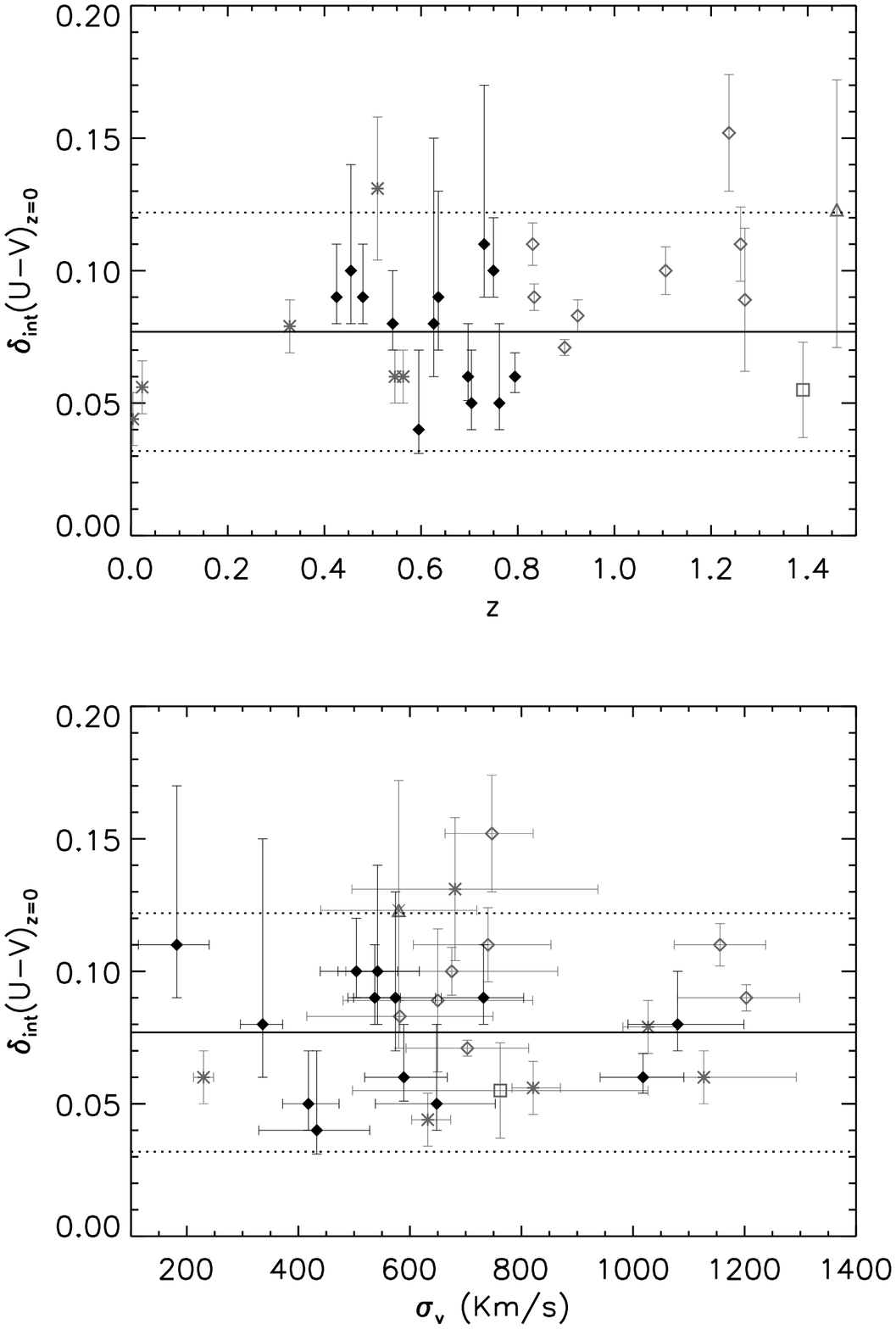}
  \end{center}   
\caption{The upper panel shows the dependence of the intrinsic scatter in the observed colour closest to rest-frame $(U-V)_{z=0}$ for EDisCS clusters (filled black diamonds) and several comparison samples at lower and higher redshift (grey symbols). The low redshift cluster sample (asterisks) was compiled from the work of \citet{BLE,vanDokkum1998} and \citet{Ellis97}. The higher redshift sample was taken from  \citet{Mei09} (open diamonds), \citet{hilton09} (open triangle) and \citet{Lidman2008} (open square).
See Table~\ref{comparison_samples} for details about the comparison samples. This plot reveals there is no significant CMR scatter evolution with redshift up to $z<1.5$. The lower panel shows the scatter as function of cluster velocity dispersion. In both panels, the solid line represents the median $\sigma_{\rm int}$ value for the EDisCS clusters and the dotted lines correspond to $\pm2\sigma$. The CMR scatter does not correlate with cluster velocity dispersion.}
\label{scatter_z}
\end{figure}

The colour scatter shows no significant evolution with redshift for the clusters in the EDisCS sample (Figure~\ref{scatter_z}, top panel). To extend the redshift baseline and compare our results with previous studies, we plot in Figure~\ref{scatter_z} similar colour scatter measurements from the sources listed in Table~\ref{comparison_samples}. No redshift dependence is found even for this extended redshift range. 
The bottom panel of Figure~\ref{scatter_z} further shows that the colour scatter does not correlate with cluster velocity dispersion ($\sigma_{v}$) either, implying that the scatter is not strongly affected by cluster mass. We note that the velocity dispersion range spanned by our sample is very broad ($200\lesssim\sigma_{v}\lesssim1200\,$km/s). Adding the clusters in the comparison samples reinforces our result. 

%where 2nd table was

\subsection{CMR scatter dependence on galaxy properties}
\label{subsec:scatter_all}

\begin{figure}
\begin{center}
  \includegraphics[width=0.5\textwidth]{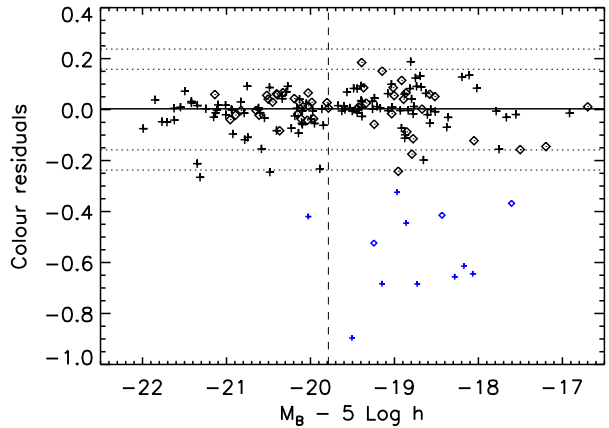}
\end{center} 
\caption{Observed colour residuals around the CMR as a function of $M_{B}$ of all the galaxies in the sample. Note that we express the magnitude as $M_{B} - 5$~Log $h$, where $h=H_{0}/70$. Elliptical galaxies are represented with crosses and lenticulars with diamonds. The solid line indicates the location of the median, dotted lines correspond to $2\sigma$ and $3\sigma$ for the black data points, and the blue symbols represent the galaxies in the ``blue tail'' (see text for details). The median luminosity is also shown for reference (black dashed line).}
 \label{res}
\end{figure}

\begin{figure}
\begin{center}
  \includegraphics[width=0.5\textwidth]{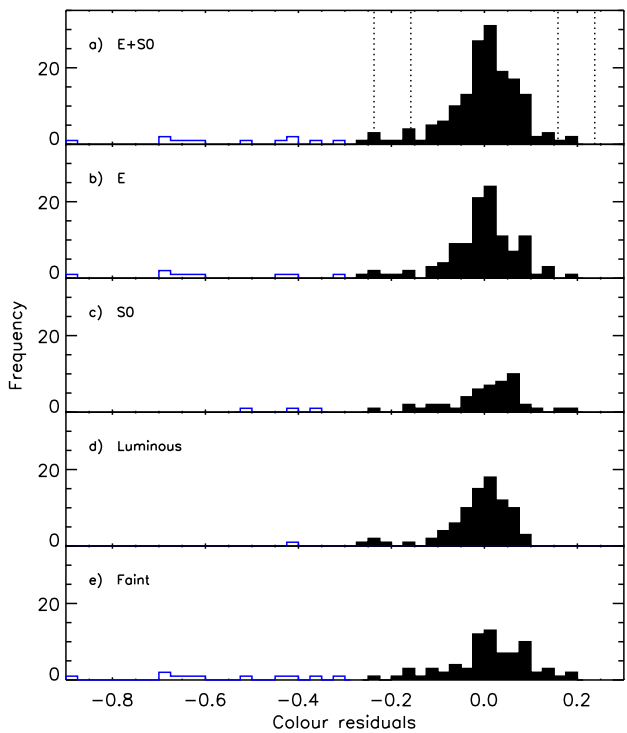}%[bb=0 0 390 140]
\end{center} 
\caption{Histograms of the observed colour residuals around the CMR for the different galaxy sub-samples computed from Figure~\ref{res}: (a) the complete sample (Ellipticals and S0s); (b) Ellipticals only (c) S0s only; (d) luminous galaxies  and (e) faint galaxies. The dotted lines in the top panel correspond to $2\sigma$ and $3\sigma$ for a Gaussian distribution with $\sigma_{\rm obs} = 0.078$. The open blue part of each histogram corresponds to the ``blue tail`` (see text and Figure~\ref{res} for details).}
 \label{histo_res}
\end{figure}

To explore the overall behavior of the galaxy colours around the CMR, Figure~\ref{res} shows the distribution of the residuals 
for the complete galaxy sample as a function of the absolute rest-frame $B$ magnitude $M_{B}$ \citep[for details of the calculation of $M_{B}$, see][]{Rudnick09}.
By construction, the residuals are concentrated about their median value ($\simeq0$; solid line in Figure~\ref{res}). The vast majority of the colour residuals follow a normal distribution reasonably well. The scatter is small, as discussed above. However, at faint magnitudes there is a clear``blue-tail'' containing a few  galaxies with significantly bluer colours (smaller blue symbols in Figure~\ref{res}).  These results can also be seen in the upper histogram of Figure~\ref{histo_res}, which shows the distribution of the colour residuals for the complete sample of early-type cluster galaxies.
The measured scatter for the whole sample is small ($\sigma_{\rm obs} = 0.078$ when excluding the ``blue-tail'') and  significantly larger than the scatter due to the photometric errors ($\simeq 0.017$). This  implies that the intrinsic colour scatter for the complete sample is $\sigma_{\rm int}=0.076^{+0.005}_{-0.004}$, very close to the average scatter for the individual clusters ($\langle\sigma_{\rm int}\rangle=0.077$, cf. Table~\ref{individual_scatters}).

\begin{table*}
\begin{center}
 \caption{Properties of the ``blue tail'' galaxies. The ``excluded'' galaxies are also listed in the two bottom lines.}
\label{outsiders}
\begin{tabular}{lclll}
\hline
EDisCS galaxy ID & Type & Residual (mag) &	$M_{B}$ & Comments (from spectra; morphology)\\
\hline\\[-2mm]
EDCSNJ1054199$-$1146065 & E	& -0.32 & -18.97	& Several emission lines	\\
EDCSNJ1054207$-$1148130 & S0	& -0.41 & -18.43	& Considerable [OII] emission \\
EDCSNJ1232307$-$1249573 & E	& -0.90 & -19.51	& Absorption-line spectra	\\
EDCSNJ1232336$-$1252103 & S0	& -0.52 & -19.24	& Galaxy of spectral type k+a	\\
EDCSNJ1232304$-$1250391 & E	& -0.42 & -20.03	& Considerable [OII] emission; Some sign of disturbance	\\
EDCSNJ1138050$-$1132546 & E	& -0.60 & -18.17	& Strong [OII] emission		\\
EDCSNJ1138068$-$1132510 & E	& -0.68 & -18.73	& Starburst?; High surface brightness	\\
EDCSNJ1138034$-$1133049 & E	& -0.64 & -18.07	& Strong [OII] emission; Some sign of disturbance\\
EDCSNJ1354073$-$1233336 & E	& -0.45 & -18.86	& Strong [OII] emission	\\
EDCSNJ1354022$-$1234283 & E	& -0.68 & -19.15	& Strong [OII] emission; Compact galaxy	\\
EDCSNJ1037564$-$1245134 & S0	& -0.37 & -17.61	& Considerable [OII] emission	\\
EDCSNJ1138135$-$1137137 & E	& -0.66 & -18.28	& Considerable [OII] emission	\\
\hline\\[-2mm]
EDCSNJ1138096$-$1135223 & $^{*}$	& -0.33 & -19.00	& Strong [OII] emission; Very disturbed/Merging	\\
EDCSNJ1138127$-$1134190 & $^{*}$	& -0.33 & -17.88	& Either HII regions or large merger	\\
\hline
\end{tabular}
\end{center}
\begin{flushleft}
	$^*$ These galaxies where misclassified as E in \citet{morph}. We excluded them from the sample since their HST images reveal 
that they are not early-type galaxies and they show strong signs of disruption  (see Section~\ref{sec:data}).
 \end{flushleft}
\end{table*}

There are 12 galaxies (7\% of the total sample) with colours $>0.3\,$mag bluer than the CMR. The number of ``blue'' galaxies in each cluster is listed in Table~\ref{cluster_sample}, while their individual IDs and some observed properties are presented in Table~\ref{outsiders}. These galaxies were excluded from our scatter-age analysis (Section \ref{sec:SFH}) for consistency with previous studies \citep[e.g.][]{Mei09} where outliers are 
rejected implicitly. We prefer to exclude them explicitly but discuss the implications that their existence and properties have in our conclusion. 
Interestingly, 2 out of the 12 blue galaxies show a small degree of disruption in their morphologies, as found by visually inspecting their HST images (see comments in Table~\ref{outsiders} and Figure~\ref{mosaic}). In Section~\ref{sec:data} we found 6 galaxies (4\%) in our full sample that, despite their clear early-type morphology, show some signs of disruption. Now we find that 2 of these disturbed galaxies have ``blue'' colours, indicating that among the blue galaxies, morphological disturbances are much more common than among the ones in the CMR. Nevertheless, the rest of the blue galaxies (10) do not show any clear sign of morphological disruption. We further discuss the implications of these faint blue galaxies (i.e. the ``blue tail'') in our conclusions.

In what follows, we concentrate on the remaining 160 galaxies whose colours are within $\pm0.3$mag from the CMR. 
From these, we constructed sub-samples according to different galaxy properties:
\begin{itemize}
 \item Morphology (E vs.\ S0), as indicated by different symbols in Figure \ref{res}.
 \item Luminosity (Luminous vs.\ Faint), divided at the median rest frame $B$ absolute luminosity (corresponding to $M_{B}^{\rm med}=-19.8$; cf. vertical dashed line in Figure~\ref{res}).
\end{itemize}
The four bottom histograms of Figure~\ref{histo_res} show the distribution of the colour residuals for each one of these sub-samples.   
Within the errors, we find that both ellipticals and S0s show the same scatter. However, the luminous galaxies have a slightly smaller intrinsic scatter than the faint ones. The values of $\sigma_{\rm int}$ for each sub-sample are listed in Table~\ref{samples_results}.
In the next section, we interpret these scatters in terms of the star-formation history of the different galaxy samples.

%where 3rd table was

%
%
%
%
%***********************************************************************************
\section{Star formation histories}
\label{sec:SFH}
\subsection{Star-formation histories of the early-type galaxies in each cluster}
\label{subsec:age}

\begin{figure}
\begin{center}
  \includegraphics[width=0.5\textwidth]{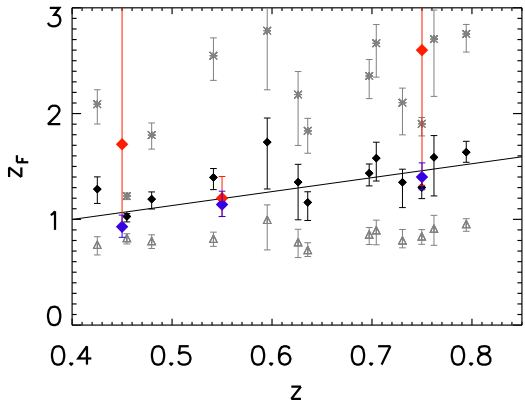}
\end{center} 
\caption{Galaxy formation redshift $z_{\rm F}$ vs.\ cluster redshift. Symbols correspond to the different values of $\beta$ used: triangles for $\beta=0.1$, filled diamonds for $\beta=0.3$ and stars for $\beta=1.0$. The solid line is a linear fit to the solid black diamonds. For comparison, the predicted $z_{\rm F}$ from \citet{SB09} are shown as larger coloured symbols. The blue diamonds correspond to the morphologically-selected sample of EDisCS early-type galaxies with galaxy velocity dispersions $< 175$km/s, while the red symbols correspond to the sample with galaxy velocity dispersions $> 175$ km/s. The blue points agree very well with our $z_{\rm F}$ for $\beta=0.3$, whilst the red points are in agreement with our results if $\beta \geq 0.3$. We observe an increase of $z_{\rm F}$ with cluster redshift (see also Figure~\ref{z_vs_zF_mei}).} 
 \label{z_vs_zF}
\end{figure}

Using the method discussed in Section~\ref{sec:method} and the intrinsic colour scatter measured for the early-type galaxies, we computed, for three values of $\beta$, the formation times $t_{\rm F}$ and their respective errors for each individual cluster or group (see table~\ref{individual_scatters}). An inspection of this table immediately shows that, as expected, for higher values of $\beta$ (less synchronous galaxy formation) older ages are required to explain the small colour scatter. A clear trend is also apparent: at a fixed $\beta$, higher redshift clusters have smaller $t_{\rm F}$. However, if we correct for the difference in look-back time using our adopted cosmology, these  $t_{\rm F}$ can be translated into formation redshifts, $z_{\rm F}$, and the trend disappears. All the EDisCS clusters yield consistent formation redshifts for their early-type galaxy population ($z_{\rm F}\simeq0.8$ for $\beta=0.1$, $z_{\rm F}\simeq1.4$ for $\beta=0.3$ and $z_{\rm F}\simeq2.4$ for $\beta=1.0$). This is shown in Figure~\ref{z_vs_zF}, where $z_{\rm F}$ is plotted vs.\ cluster redshift for the different values on $\beta$. For comparison, we overplot the formation redshifts derived by \citet{SB09} using absorption-line indices in the spectra of EDisCS early-type red-sequence galaxies (blue and red diamonds).
The blue points (corresponding to galaxies with velocity dispersions $< 175$km/s) agree very well with our formation redshifts for $\beta=0.3$ (black diamonds), while the red points (galaxy velocity dispersions $> 175$km/s) agree with $\beta\geq0.3$.
Moreover, galaxy ages derived from the analysis of the Fundamental Plane of these galaxies (Saglia et al.\ in preparation) are also in agreement with our formation redshifts for $\beta\geq0.3$.

\begin{figure}
\begin{center}
  \includegraphics[width=0.5\textwidth]{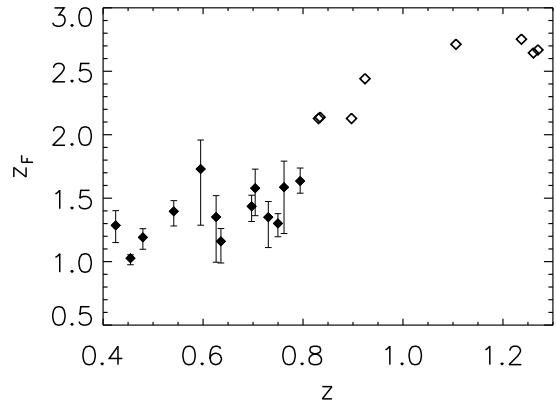}
\end{center} 
\caption{Galaxy formation redshift $z_{\rm F}$ (derived using $\beta=0.3$) vs.\ cluster redshift for EDisCS clusters (solid diamonds) and for the clusters published by \citet{Mei09} (open diamonds). We observe that $z_{\rm F}$ increases slightly with cluster redshift for the EDisCS sample (see also fitted line in Figure~\ref{z_vs_zF}). This trend becomes stronger when we include the higher redshift comparison sample.}
 \label{z_vs_zF_mei}
\end{figure}

Figure~\ref{z_vs_zF} also shows that $z_{\rm F}$ may be slightly higher for higher redshift EDisCS clusters (fitted line). Although this trend is not very significant, it would be desirable to extend the redshift baseline to test whether it continues at higher redshifts. We can do that by using the study published by \citet{Mei09}. These authors follow a very similar procedure to ours, and predict formation times based on a colour-scatter analysis for a value of $\beta\simeq0.3$. Their galaxy samples also contain morphologically-classified ellipticals and S0s, and can therefore be compared to ours. Their results are plotted as open diamonds in Figure~\ref{z_vs_zF_mei}, together with our results for $\beta=0.3$. If we take these points at face value, the trend of increasing $z_{\rm F}$ with redshift becomes very significant. However, a word of caution is required. Although our study and that of \citet{Mei09} are very similar, there are some differences. First, their photometry and the colours that they use are different because the higher redshift of their clusters. Second, they estimate the intrinsic colour scatter using biweight scale estimator (which implicitly excludes outliers). However, using their method with our data does not change our results since we exclude outliers explicitly, as discussed in section~\ref{sec:method}. Finally, their implementation of the scatter-age test differs in some minor details from ours, although they use the same \citet{bc03} models. We believe these differences are probably not important for this exercise, but we cannot be completely certain without re-analysing their data. With all these caveats, the trend observed in Figure~\ref{z_vs_zF_mei} would imply that that morphologically-classified elliptical and S0 galaxies formed earlier in higher redshift clusters than in lower redshift ones.

\begin{figure}
\begin{center}
  \includegraphics[width=0.5\textwidth]{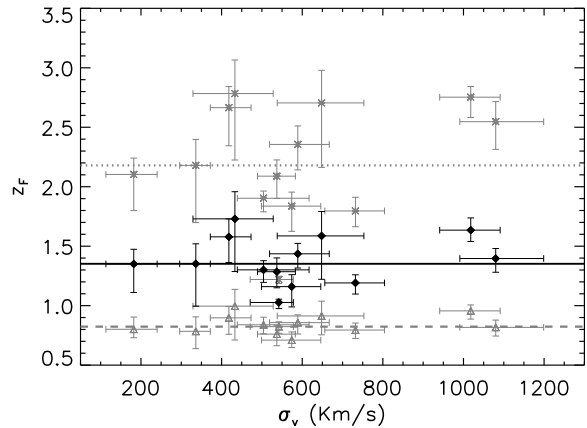}
\end{center} 
\caption{Galaxy formation redshift $z_{\rm F}$ vs.\ cluster velocity dispersion. As in Figure \ref{z_vs_zF}, symbols correspond to the different values of $\beta$: triangles for $\beta=0.1$, filled diamonds for $\beta=0.3$ and stars for $\beta=1.0$. The dashed, solid and dotted lines correspond to the median formation redshift
for each value of $\beta$. We find that $z_{\rm F}$ does not depend on the cluster velocity dispersion.}
 \label{sigma_vs_zF_ediscs}
\end{figure}

Figure~\ref{sigma_vs_zF_ediscs} shows $z_{\rm F}$ plotted against cluster velocity dispersion. Consistent with Figure~\ref{scatter_z}, no correlation is found between $z_{\rm F}$ and $\sigma_{v}$, implying that the formation time of morphologically-classified ellipticals and S0s does not depend strongly on cluster mass.

\subsection{Dependence of the star-formation histories on galaxy properties}
\label{subsec:age_all}

\begin{table*}
\begin{center}
\caption{Scatter analysis results for the morphology- and luminosity-split sub-samples. The colour scatter and the derived formation times and  redshifts are given for three values of $\beta$. }
\label{samples_results}
\begin{tabular}{lccllllll}
\hline \\[-2mm]
\textbf{Sample}	&$\langle z\rangle$	& $\sigma_{\rm int}$ 	&$t_{\rm F}(\beta=0.1)$&	$t_{\rm F}(\beta=0.3)$&$t_{\rm F}(\beta=1.0)$ \qquad\qquad & $z_{\rm F}(\beta=0.1)$ 	&$z_{\rm F}(\beta=0.3)$ 	& $z_{\rm F}(\beta=1.0)$ \\
\hline \\[-2mm]

All	&$0.6411$	&$0.077^{+0.005}_{-0.004}$ &$2.64^{+0.07}_{-0.08}$	&$4.04^{+0.06}_{-0.07}$	&$5.45^{+0.06}_{-0.07}$	&	$1.22^{+0.3}_{-0.3}$&		$1.85^{+0.04}_{-0.05}$	&$2.88^{+0.07}_{-0.08}$	\\[1mm]
\hline \\[-2mm]
E	&$0.6443$	&$0.075^{+0.005}_{-0.004}$&$2.65^{+0.09}_{-0.1}$	&$4.05^{+0.08}_{-0.09}$	&$5.45^{+0.07}_{-0.09}$	&	$1.22^{+0.3}_{-0.3}$&		$1.85^{+0.04}_{-0.05}$	&$2.88^{+0.07}_{-0.08}$	\\[1mm]
S0	&$0.6589$	&$0.079^{+0.009}_{-0.007}$&$2.6^{+0.1}_{-0.2}$		&$4.0^{+0.1}_{-0.1}$		&$5.3^{+0.1}_{-0.1}$		&	$1.24^{+0.05}_{-0.05}$&		$1.87^{+0.06}_{-0.07}$	&$2.9^{+0.1}_{-0.1}$ 		\\[1mm]
\hline \\[-2mm]
Luminous&$0.7016$	&$0.070^{+0.006}_{-0.005}$&$2.636\pm0.1$		&$4.00^{+0.08}_{-0.1}$	&$5.33^{+0.08}_{-0.09}$	&	$1.34^{+0.04}_{-0.05}$&	$2.03^{+0.08}_{-0.09}$	&$3.1^{+0.2}_{-0.2}$	\\[1mm]
Faint	&$0.6071 $	&$0.080^{+0.007}_{-0.006}$&$2.668^{+0.1}_{-0.1}$		&$4.10^{+0.09}_{-0.1}$	&$5.6^{+0.1}_{-0.1}$		&	$1.11^{+0.03}_{-0.04}$&		$1.67^{+0.05}_{-0.05}$		&$2.65^{+0.08}_{-0.09}$	\\[1mm]
\hline \\[-2mm]
\end{tabular}
\end{center}
\end{table*}

In table~\ref{samples_results}, we show the colour scatter and derived formation redshift for the galaxy samples divided in terms of morphology and luminosity, as discussed in Section~\ref{subsec:scatter_all}. Within the errors, ellipticals and S0s show the same scatter.
Since the E and S0 samples have very similar mean redshifts, similar colour scatters imply similar average formation redshift.
However, the faint galaxies seem to exhibit a larger scatter than the bright ones. This, together with the fact that the luminous subsample has a higher average redshift than the faint one, suggests that the most luminous (massive) early-type galaxies formed earlier than the fainter (less massive) ones. 

An identical conclusion is reached if the sample is split by stellar mass \citep[derived following][]{BelldeJong01} 
instead of by luminosity, which is not surprising since homogeneous colours imply near-constant stellar mass-to-light ratios. When splitting the sample by galaxy velocity dispersion or dynamical mass (cf. Saglia et al.\ in preparation), similar trends are observed, albeit with smaller statistical significance since only half of the galaxies in our sample have measured velocity dispersions.

%
%****************************************************************************************
\section{The CMR zeropoint}
\label{sec:zeropoint}

\begin{figure}
\begin{center}
  \includegraphics[width=0.5\textwidth]{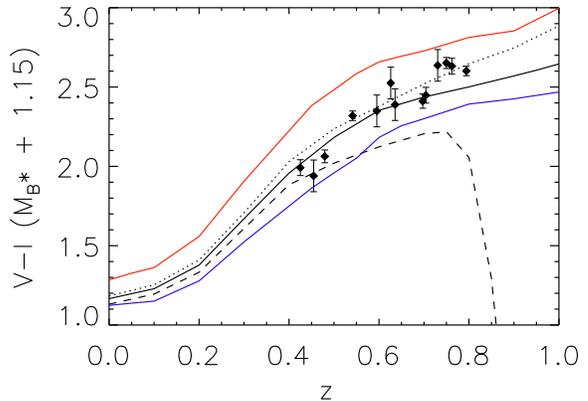}
\end{center} 
\caption{ $V-I$ CMR zero-point vs.\ redshift  for the EDisCS cluster galaxies (points) compared with population synthesis models from \citet{bc03}. 
The black lines have been computed with $Z_{\rm solar}=0.02$ metallicity, the blue one with $Z_{\rm sub-solar}=0.008$, and the red one with $Z_{\rm super-solar}=0.05$. The black dashed line corresponds to $\beta=0.1$ ($z_{\rm F}\sim0.9$), the black, blue and red solid lines to $\beta=0.3$ ($z_{\rm F}\sim1.5$) and the black dotted line to $\beta=1.0$ ($z_{\rm F}\sim2.5$). See text for details. The observed points clearly rule out a $\beta=0.1$ scenario (i.e. very synchronized formation for all the galaxies). However, they are in agreement with $\beta \geq 0.3$.}
 \label{col_z}
\end{figure}

Our CMR scatter analysis cannot constrain $t_{\rm F}$ and $\beta$ simultaneously. Until now, we have not discussed the zero-point of the CMR because predicting absolute colours from stellar population models is, arguably, more uncertain than predicting differential colour changes \citep[see, e.g.,][]{Aragon1993}. However, since accurate zero points are available, it is worthwhile checking whether consistent and, perhaps, additional constraints can be obtained from them. Figure~\ref{col_z} shows for each of the EDisCS clusters the CMR colour corresponding to the median absolute $B$ magnitude of the sample, after correcting for luminosity evolution. Specifically, for each cluster we determine the colour of its CMR for $M_{\rm B}=M_{\rm B}^*+1.15$, where $M_{\rm B}^*$ is empirically-determined \citep{Crawford2009,Rudnick09}. These zero-points do not correlate with intrinsic cluster properties such as their velocity dispersions or masses.

The solid lines in the Figure show the predictions of \citet{bc03} models for $z_{\rm F}\sim1.5$ (corresponding to $\beta=0.3$, cf.\ Section~\ref{subsec:age}) for $Z_{\rm sub-solar}=0.008$ (blue), $Z_{\rm solar}=0.02$ (black) and $Z_{\rm super-solar}=0.05$ (red). It is clear that for median luminosity galaxies non-solar models do not provide an acceptable fit to the galaxy colours for any value of $\beta$.\footnote{For clarity, we show the non-solar metallicity lines for $\beta=0.3$ only.} However, solar-metallicity models do a reasonable job. This provides additional justification for the use of solar metallicity models in our analysis (cf.\ section~\ref{sec:method}). 

Taking the solar models at face value, a constraint on $\beta$ can be derived from Figure~\ref{col_z}. The black-dashed line corresponds to $\beta=0.1$ ($z_{\rm F}\sim0.9$), the black solid line to $\beta=0.3$ ($z_{\rm F}\sim1.5$) and the black dotted line to $\beta=1.0$ ($z_{\rm F}\sim2.5$). The observed points clearly rule out $\beta=0.1$ (i.e. very synchronized formation for all the galaxies). They are in reasonably good agreement with $\beta=1.0$, but $\beta=0.3$ is not ruled out. Hence, it seems reasonably safe to conclude that $\beta \geq 0.3$ on the basis of this analysis. If we translate this into the time interval  $\Delta t$ over which all the galaxies ``formed'', the constraint translates to $\Delta t \gtrsim 1\,$Gyr. Since $t_{\rm F}$ refers to the time at which star formation ceased, this implies that there was an extended epoch over which cluster galaxies had their star formation truncated/stopped. In other words, this cessation of star formation was not synchronized for all the cluster early-type galaxies.

%
%****************************************************************************************
\section{Discussion and Conclusions}
\label{sec:conclusions}

In this paper, we have studied the colour-magnitude relation (CMR) for a sample of early-type galaxies from the ESO Distant Cluster Survey (EDisCS). Our sample consists of 172 strictly morphologically-classified ellipticals and S0 galaxies in 13 clusters and groups with redshifts $0.4<z<0.8$ and velocity dispersions $200\lesssim \sigma_v\lesssim1200\,$km/s. All these galaxies are spectroscopically-confirmed cluster members, and their magnitudes span the range $-22\lesssim M_B - 5\log h \lesssim -17.5$. We have analyzed the colour scatter about the CMR and its zeropoint to derive meaningful constraints on the formation history of these galaxies. Assuming that the intrinsic colour scatter about the CMR is due to differences in stellar population ages,
our main results are:

\begin{itemize}

\item In agreement with previous studies, the intrinsic colour scatter $\sigma_{\rm int}$ about the CMR in rest-frame $U-V$ is small 
($\langle\sigma_{\rm int} \rangle= 0.076$). However, there is a small minority of faint early-type galaxies (7\%) that are significantly bluer than the CMR and where excluded from the scatter analysis. These galaxies probably represent a population of young galaxies that have not yet joined the red-sequence population. Interestingly, only 2 out of the 12 blue galaxies show signs of morphological disturbances and/or interactions, while the rest are bona-fide ellipticals or S0s. However, the vast majority of the blue galaxies have emission lines in their spectra indicative of ongoing star formation (see table~\ref{outsiders}). Faint blue low-mass early-type galaxies have been reported in previous studies  \citep[e.g.][]{Blakeslee2006,Bamford2009}, preferentially in low density environments. To explain the existence of these low-mass blue early-type galaxies with normal E/S0 morphologies
in the field \citet{Huertas10} propose two possible explanations. First, they could be the result of minor mergers, which would trigger centrally-concentrated star-formation, helping to build a bulge, and eventually taking them to the red sequence. Alternatively, the disks in these galaxies are perhaps being (re)built from the surrounding gas, moving then back (or staying) in the blue-cloud. It is hard to see how this second possibility would work in the cluster environment, where it is more likely that gas is removed than allowed to fall onto these low mass galaxies. In clusters, minor mergers remain a possibility, in particular if they occur while these galaxies were in filaments and/or groups, but they need to be minor enough to avoid strong morphological disruption. It is also possible that these galaxies are just approaching the cluster for the first time, and will eventually stop forming stars due to gas removal by the cluster environment. This would take them to the red sequence 
without severely disrupting their morphologies.

\item We observe no significant evolution of the intrinsic colour scatter to $z\simeq0.8$ from the EDisCS clusters alone. This result is consistent with previous studies \citep[e.g.][]{Ellis97}. After expanding our sample with higher redshift clusters from the literature, we have still found no significant evolution in $\sigma_{\rm int}$ up to $z\sim1.5$. Moreover, in the wide range of cluster velocity dispersion (mass) of our sample ($100\lesssim \sigma_{v} \lesssim 1300$) the scatter does not seem to show any trend. Because our sample is strictly morphologically-selected, this implies that by the time cluster elliptical and S0 galaxies achieve their morphology, the vast majority have already joined the red sequence. The only exception seems to be the very small fraction ($\lesssim7\%$) of faint blue early-types. 

\item Following the work of \citet{BLE}, we used the colour scatter to estimate the galaxies' formation time $t_{\rm F}$, defined as the time elapsed since the major episode of star formation. This allowed us to calculate the formation redshift $z_{\rm F}$ for the early-type galaxy population in each cluster. Yet again, we measured no significant dependency of $z_{\rm F}$ on the cluster velocity dispersion. However, we found that $z_{\rm F}$ increases weakly with cluster redshift within the EDisCS sample. This trend becomes very clear when the higher redshift clusters from \citet{Mei09} are included. This implies that, at any given redshift, to have a population of fully-formed ellipticals and S0s they must have formed most of their stars $\simeq2$--$4\,$Gyr prior to observation. That does not mean that \textit{all} early-type galaxies in \textit{all} clusters formed at these high redshifts. It means that the ones that we observe to already have early-type morphologies also have reasonably old stellar populations. This is partly a manifestation of the ``progenitor bias'' \citep{vanDokkum1996}, but also a consequence of the vast majority of the early-type galaxies in clusters (in particular the massive ones) being already red (i.e., already having old stellar populations) by the time the achieved their morphology. 

\item Elliptical and S0 galaxies show very similar colour scatter, implying that they have similar stellar population ages. If we assume that their observed properties are representative of the early-type cluster galaxy population at these redshifts, the scarcity of blue S0s indicates that, if they are the descendants of spirals whose star formation has ceased \citep{Aragon2006,Bedregal2006,Barr2007}, the galaxies were already red when they became S0s, i.e.\ the parent spiral galaxies became red before loosing their spiral arms. The red spirals found preferentially in dense environments \citep{Wolf2009,Bamford2009, Masters2009} are the obvious candidate progenitors of these S0s. 

\item Dividing the sample in two halfs by luminosity (or stellar mass), we find that the formation redshift $z_{\rm F}$ (derived from the CMR scatter in each sample) is smaller for fainter galaxies than for brighter ones. This indicates that fainter early-type galaxies finished forming their stars later. Our results are also consistent with the observation that the cluster red sequence built over time with the brightest galaxies reaching the sequence earlier than fainter ones \citep{DeLucia04,DeLucia07,Rudnick09}.

\item The CMR scatter analysis cannot constrain both the formation time $t_{\rm F}$ and formation interval $\Delta t$ simultaneously. However, its combination with the observed evolution of the CMR zero point, enabled us to conclude that the early-type cluster galaxy population must have had 
their star formation truncated/stopped over an extended period $\Delta t \gtrsim 1\,$Gyr. Hence, the cessation of star formation was not synchronized for all the cluster early-type galaxies.

\end{itemize}

\section*{Acknowledgments}
Based on observations collected at the European Southern Observatory, Chile, as part of programme 073.A-0216. YLJ acknowledges the European Southern Observatory for financial support and hospitality during the writing of this paper. We also thank 
Claire Halliday, Bo Milvang-Jensen, Bianca Poggianti and Piero Rosati for useful discussions. GDL acknowledges financial support from the European Research Council under the European Community's Seventh Framework Programme (FP7/2007-2013)/ERC grant agreement n. 202781

\bsp

%%%%%%%%%%%%%%%%%%%%%%%%%%%%%%%%%%%%%%%%%%%%%%%%%%%%%%%%%%%%%%%%%%%%%%%%%%%%%%%%%%%%%%%%%%%%%%%%%%%%%
%
%			T   A   B   L   E   S
%
%%%%%%%%%%%%%%%%%%%%%%%%%%%%%%%%%%%%%%%%%%%%%%%%%%%%%%%%%%%%%%%%%%%%%%%%%%%%%%%%%%%%%%%%%%%%%%%%%%%%%

\appendix

%%%%%%%%%%%%%%%%%%%%%%%%%%%%%%%%%%%%%%%%%%%%%%%%%%%%%%%%%%%%%%%%%%%%%%%%%%%%%%%%%%%%%%%%%%%%%%%%%%%%%

%%%%%%%%%%%%%%%%%%%%%%%%%%%%%%%%%%%%%%%%%%%%%%%%%%%%%%%%%%%%%%%%%%%%%%%%%%%%%%%%%%%%%%%%%%%%%%%%%%%%%]

\end{document}